\documentclass[12pt]{article}
\begin{document}

\title{Five-loop renormalization-group expansions for 
two-dimensional Euclidean $\lambda \phi^4$ theory}
\author{E. V. Orlov, ~A. I. Sokolov \\
\em {Saint Petersburg Electrotechnical University,} \\ 
\em {197376, Saint Petersburg, Russia}}
\date{May 19, 2000}
\maketitle

\begin{abstract}
{The renormalization-group functions of the two-dimensional 
$n$-vector $\lambda \phi^4$ model are calculated in the five-loop 
approximation. Perturbative series for the $\beta$-function and 
critical exponents are resummed by the Pade-Borel-Leroy techniques. 
An account for the five-loop term shifts the Wilson fixed point 
location only briefly, leaving it outside the segment formed by 
the results of the lattice calculations. This is argued to reflect 
the influence of the non-analytical contribution to the 
$\beta$-function. The evaluation of the critical exponents for 
$n = 1$, $n = 0$ and $n = -1$ in the five-loop approximation and 
comparison of the results with known exact values confirm the 
conclusion that non-analytical contributions are visible in two 
dimensions. For the $2D$ Ising model, the estimate 
$\omega = 1.31(3)$ for the correction-to-scaling exponent is 
found that is close to the values resulting from the 
high-temperature expansions.}    
\end{abstract}

The field-theoretical renormalization-group (RG) approach proved 
to be a powerful tool for calculating the critical exponents and 
other universal quantities of the basic three-dimensional (3D) 
models of phase transitions. Today, many-loop RG expansions for 
$\beta$-functions (six-loop), critical exponents (seven-loop), 
higher-order couplings (four-loop), etc. of the 3D $O(n)$-symmetric 
and cubic models are known resulting in the high-precision numerical 
estimates for experimentally accessible quantities [1-7]. The main 
aim of this report is to elaborate further the field-theoretical RG 
technique, namely, to clear up how effective is this machinery in 
two dimensions where i) the RG series are stronger divergent and 
ii) singular (non-analytic) contributions to RG functions are 
expected to be larger than for 3D systems.  

The Hamiltonian of the model describing the critical behaviour of 
various two-dimensional systems reads:
\begin{equation}
H = 
\int d^2x \Biggl[{1 \over 2}( m_0^2 \varphi_{\alpha}^2
 + (\nabla \varphi_{\alpha})^2) 
+ {\lambda \over 24} (\varphi_{\alpha}^2)^2 \Biggr] ,
\label{eq:1} \\
\end{equation}
where $\varphi_{\alpha}$ is a real $n$-vector field, $m_0^2$ is 
proportional to $T - T_c^{(0)}$, $T_c^{(0)}$ being the mean-field 
transition temperature. 

We calculate the $\beta$-function and the critical exponents for 
the model (1) within the massive theory. The Green function, the 
four-point vertex and the $\phi^2$ insertion are normalized in a 
conventional way:
\begin{eqnarray}
G_R^{-1} (0, m, g_4) = m^2 , \qquad \quad 
{{\partial G_R^{-1} (p, m, g_4)} \over {\partial p^2}}
\bigg\arrowvert_{p^2 = 0} = 1 , \\
\nonumber
\Gamma_R (0, 0, 0, m, g) = m^2 g_4, \qquad \quad  
\Gamma_R^{1,2} (0, 0, m, g_4) = 1. 
\label{eq:2}
\end{eqnarray}

Since the four-loop RG expansions at $n = 1$ are known [1] we are 
in a position to find corresponding series for arbitrary $n$ and 
to calculate the five-loop terms. The results of our calculations 
are as follows:    
\begin{eqnarray}
{\beta (g) \over 2} = - g + g^2 - {g^3 \over (n + 8)^2} 
\biggl( 10.33501055 n + 47.67505273 \biggr) \ \qquad \qquad \qquad \qquad
\nonumber \\
+ {g^4 \over (n + 8)^3} \biggl( 5.000275928 n^2 
+ 149.1518586 n + 524.3766023 \biggr) \ \qquad \qquad \qquad \qquad
\nonumber \\
- {g^5 \over (n + 8)^4} \biggl( 0.0888429 n^3 + 179.69759 n^2   
+ 2611.1548 n + 7591.1087 \biggr) \ \ \qquad \qquad 
\nonumber \\
+ {g^6 \over (n + 8)^5} 
\biggl( -0.00408 n^4 + 80.3096 n^3   
+ 5253.56 n^2 + 53218.6 n + 133972 \biggr). \qquad   
\label{eq:3} 
\end{eqnarray}
  
\begin{eqnarray}
\gamma^{-1} = 1 - {{n + 2} \over {n + 8}}~g 
+ {g^2 \over (n + 8)^2}~(n + 2)~3.375628955 
\qquad \qquad \qquad \qquad \qquad \qquad
\nonumber \\
- {g^3 \over (n + 8)^3} \biggl( 4.661884772 n^2 + 34.41848329 n 
+ 50.18942749 \biggr) \ \qquad \qquad \qquad \qquad 
\nonumber \\
+ {g^4 \over (n + 8)^4} \biggl( 0.3189930 n^3 + 71.703302 n^2 
+ 429.42449 n + 574.58772 \biggr) \ \ \qquad \qquad
\nonumber \\
- {g^5 \over (n + 8)^5} \biggl( 0.09381 n^4 + 85.4975 n^3 
+ 1812.19 n^2 + 8453.70 n + 10341.1 \biggr) . \qquad
\label{eq.4}
\end{eqnarray}
\begin{eqnarray}
\eta = {g^2 \over (n + 8)^2}~(n + 2)~0.9170859698 
- {g^3 \over (n + 8)^2}~(n + 2)~0.05460897758 \qquad \quad
\nonumber \\
+ {g^4 \over (n + 8)^4} \biggl( - 0.09268446 n^3 + 4.0564105 n^2 
+ 29.251167 n + 41.535216 \biggr) \ \qquad
\nonumber \\
- {g^5 \over (n + 8)^5} \biggl( 0.07092 n^4 + 1.05240 n^3 
+ 57.7615 n^2 + 325.329 n + 426.896 \biggr) . \qquad
\label{eq.5}
\end{eqnarray}
In more details, the calculations are described in Ref.[8]. Instead 
of the renormalized coupling constant $g_4$, a rescaled coupling  
\begin{equation}
g = {n + 8 \over {24 \pi}} g_4,
\label{eq:6}
\end{equation}
is used as an argument in above RG series. This variable is more 
convenient since it does not go to zero under $n \to {\infty}$ but 
approaches the finite value equal to unity.

To evaluate the Wilson fixed point location $g^*$ and numerical 
values of the critical exponents, the resummation procedure based 
on the Borel-Leroy transformation 
\begin{equation}
f(x) = \sum_{i = 0}^{\infty} c_i x^i = \int\limits_0^{\infty} 
e^{-t} t^b F(xt) dt, \ \ \ \ \   
\nonumber\\
F(y) = \sum_{i = 0}^{\infty} {c_i \over (i + b)!} y^i \ \ , 
\label{eq:7}
\end{equation}
is used. The analytical extension of the Borel transforms is 
performed by exploiting relevant Pad\'e approximants [L/M]. In 
particular, four subsequent diagonal and near-diagonal approximants 
$[1/1]$, $[2/1]$, $[2/2]$, and $[3/2]$ turn out to lead to numerical 
estimates for $g^*$ which rapidly converge, via damped oscillations, 
to the asymptotic values; this is cleary seen from Table 1. These 
asymptotic values, i. e. the final five-loop RG estimates for $g^*$ 
are presented in Table 2 for $0 \le n \le 8$ (to avoid confusions, 
let us note that models with $n \ge 2$ possessing no ordered phase 
are studied here only as polygons for testing the numerical power of 
the perturbative RG technique). As Table 2 demonstrates, the numbers 
obtained differ appreciably from numerical estimates for $g^*$ given 
by the lattice and Monte Carlo calculations [9-15]; such estimates 
are usually extracted from the data obtained for the linear ($\chi$) 
and non-linear ($\chi_4$) susceptibilities related to each another 
via $g_4$: 
\begin{equation}
\chi_4 = {\partial^3M \over{\partial H^3}} \Bigg\arrowvert_{H = 0} 
= - \chi^2 m^{-2} g_4, \qquad \quad 
\label{eq:8} \\
\end{equation}
Since the convergence of the numerical estimates for $g^*$ given 
by the resummed RG series is oscillatory, an 
account for higher-order (six-loop, seven-loop, etc.) terms in the 
expansion (3) will not avoid this discrepancy [8]. That is why we 
believe that it reflects the influence of the singular 
(non-analytical) contribution to the $\beta$-function. 

The critical exponents for the Ising model ($n = 1$) and for those 
with $n = 0$ and $n = -1$ are estimated by the Pad\'e-Borel summation 
of the five-loop expansions (4), (5) for $\gamma^{-1}$ ¨ $\eta$. Both 
the five-loop RG (Table 1) and the lattice (Table 2) estimates for 
$g^*$ are used in the course of the critical exponent evaluation. 
To get an idea about an accuracy of the numerical results obtained 
the exponents are estimated using different Pad\'e approximants, 
under various values of the shift parameter $b$, etc. In particular, 
the exponent $\eta$ is estimates in two principally different ways: 
by direct summation of the series (5) and via the resummation of RG 
expansions for exponents
\begin{equation}
\eta^{(2)} = {1 \over \nu} + \eta - 2, 
\qquad \qquad \eta^{(4)} = {1 \over \nu} - 2,  
\label{eq:9}
\end{equation}
which possess a regular structure favouring the rapid convergence of 
the iteration procedure. The typical error bar thus found is about 
0.05. 

The results obtained are collected in Table 3. As is seen, for 
small exponent $\eta$ and in some other cases the differences between 
the five-loop RG estimates and known exact values of the critical 
exponents exceed the error bar mentioned. Moreover, in the five-loop 
approximation the correction-to-scaling exponent $\omega$ of the 2D 
Ising model is found to be close to the value 4/3 predicted by the 
conformal theory [16] and to the estimate $1.35 \pm 0.25$ extracted 
from the high-temperature expansions [17] but differs markedly from 
the exact value $\omega = 1$ [18] and contradicts to recent 
conjecture $\omega = 2$ [19]. This may be considered as an argument 
in favour of the conclusion that non-analytical contributions are 
visible in two dimensions.  

We thank B. N. Shalaev for numerous useful discussions of the 
critical thermodynamics of 2D systems. The work was supported by the 
Ministry of Education of Russian Federation (Grant 97-14.2-16), by 
the International Science Foundation (A. I. S., Grant à99-943), and 
by Saint Petersburg Administration (E. V. O., grant ASP 298496).

\newpage
\begin{table}
\caption{The Wilson fixed point coordinate for models with $n = 1$, 
$n = 0$ and $n = -1$ in four subsequent RG approximations and the 
final five-loop estimates for $g^*(n)$.}
\begin{tabular}{|c|c|c|c|c|c|}
\hline
~~~$n$~~~& [1/1]~~~& [2/1]~~~& [2/2]~~~& [3/2]~~~& $g^*$, 5-loop~~\\
\hline
\hline
1  & 2.4246~~~~& 1.7508~~~~& 1.8453~~~~& 1.8286~~~~& 1.837 $\pm$ 
0.03~~~~\\
\hline
0  & 2.5431~~~~& 1.7587~~~~& 1.8743~~~~& 1.8402~~~~& 1.86 $\pm$ 
0.04~~~~\\
\hline
-1 & 2.6178~~~~& 1.7353~~~~& 1.8758~~~~& 1.8278~~~~& 1.85 $\pm$ 
0.05~~~~\\
\hline
\end{tabular}
\end{table}

\begin{table}
\caption{The Wilson fixed point coordinate $g^*$ and critical exponent  
$\omega$ for $0 \le n \le 8$ obtained in the five-loop RG 
approximation. The values of $g^*$ extracted from high-temperature (HT) 
[10,12] and strong coupling (SC) [11] expansions, found by Monte Carlo 
simulations (MC) [13,15], obtained by the constrained resummation of 
the $\epsilon$-expansion for $g^*$ ($\epsilon$-exp.) [12], and given 
by corresponding $1/n$-expansion ($1/n$-exp.) [12] are also presented 
for comparison.}
\begin{tabular}{|c|c|c|c|c|c|c|}
\hline
$n$ & 0 & 1 & 2 & 3 & 4 & 8 \\
\hline
\multicolumn{7}{|c|}{$g^*$} \\
\hline
RG, 5-loop & 1.86(4) & 1.837(30) & 1.80(3) & 1.75(2) 
& 1.70(2) & 1.52(1) \\
&  &  &  &  & ($b=1$) & ($b=1$) \\
\hline
HT & 1.679(3) & 1.754(1) & 1.81(1) & 1.724(9) 
& 1.655(16) &  \\
\hline
MC &  & 1.71(12) & 1.76(3) & 1.73(3) &  &  \\  
\hline
SC & 1.673(8) & 1.746(8) & 1.81(2) & 1.73(4) &  &  \\
\hline
$\epsilon$-exp. & 1.69(7) & 1.75(5) & 1.79(3) & 1.72(2) 
& 1.64(2) & 1.45(2) \\
\hline
1/n-exp. &  &  &  & 1.758 & 1.698 & 1.479 \\
\hline
\hline
\multicolumn{7}{|c|}{$\omega$} \\
\hline
RG, 5-loop & 1.31(3) & 1.31(3) & 1.32(3) & 1.33(2) 
& 1.37(3) & 1.50(2) \\
\hline
\end{tabular}
\end{table}

\begin{table}
\caption{Critical exponents for $n = 1$, $n = 0$, and $n = -1$ 
obtained via the Pad\'e-Borel summation of the five-loop RG 
expansions for $\gamma^{-1}$ and $\eta$. The known exact values 
of these exponents are presented for comparison.}
\begin{tabular}{|c|c|c|c|c|c|c|}
\hline
~~~$n$~~~&  & $g^*$ & $\gamma$ & $\eta$ & $\nu$ & $\alpha$ \\
\hline
\hline
1  & RG    & 1.837        & 1.790 & 0.146 & 0.966~~~& 0.068~~~ \\
   &       & 1.754 (HT)~~~& 1.739 & 0.131 & 0.931~~~& 0.139~~~ \\
\hline   
   & exact~~&              &  7/4   &  1/4   &  1     &  0     \\   
   &       &              & (1.75) & (0.25) &        &         \\
\hline   
\hline
0  & RG    & 1.86         & 1.449 & 0.128 & 0.774 & 0.452 \\
   &       & 1.679 (HT)~~~& 1.402 & 0.101 & 0.738 & 0.524 \\
\hline
   & exact~~& & 43/32        & 5/24         & 3/4~~  & 1/2~~ \\
   &       & & (1.34375)~~~~& (0.20833)~~~~& (0.75) & (0.5) \\
\hline
\hline
-1 & RG    & 1.85         & 1.184 & 0.082 & 0.617 & 0.765 \\
   &       & 1.473 (SC)~~~& 1.155 & 0.049 & 0.592 & 0.816 \\
\hline   
   & exact~~& & 37/32        &  3/20  &   5/8   &  3/4   \\
   &       & & (1.15625)~~~~& (0.15) & (0.625) & (0.75)  \\
\hline
\end{tabular}
\end{table}
\end{document}